\documentclass[twocolumn,showpacs,amsmath,amssymb,pra,superscriptaddress,floatfix]{revtex4}

\usepackage{amssymb}
\usepackage{graphicx}
\usepackage{dcolumn}
\usepackage{bm}
\usepackage{psfrag}

\begin{document}

\title{Creating collective many-body states with highly excited atoms}

\author{B. Olmos}
\email{bolmos@ugr.es}
\affiliation{Instituto 'Carlos I' de F\'{\i}sica Te\'orica y Computacional
and Departamento de F\'{\i}sica At\'omica, Molecular y Nuclear,
Universidad de Granada, E-18071 Granada, Spain}
\affiliation{Midlands Ultracold Atom Research Centre - MUARC, The University of Nottingham, School of Physics and Astronomy, Nottingham, United Kingdom}
\author{R. Gonz\'{a}lez-F\'{e}rez}
\email{rogonzal@ugr.es}
\affiliation{Instituto 'Carlos I' de F\'{\i}sica Te\'orica y Computacional
and Departamento de F\'{\i}sica At\'omica, Molecular y Nuclear,
Universidad de Granada, E-18071 Granada, Spain}
\author{I. Lesanovsky}
\email{igor.lesanovsky@nottingham.ac.uk}
\affiliation{Midlands Ultracold Atom Research Centre - MUARC, The University of Nottingham, School of Physics and Astronomy, Nottingham, United Kingdom}

\date{\today}
\begin{abstract}
We study the collective excitation of a gas of highly excited atoms confined to a large spacing ring lattice, where the ground and the excited states are coupled resonantly via a laser field. Our attention is focused on the regime where the interaction between the highly excited atoms is very weak in comparison to the Rabi frequency of the laser. We demonstrate that in this case the many-body excitations of the system can be expressed in terms of free spinless fermions. The complex many-particle states arising in this regime are characterized and their properties, e.g. their correlation functions, are studied. In addition we investigate how one can actually experimentally  access some of these many-particle states by a temporal variation of the laser parameters.
\end{abstract} \maketitle

\section{Introduction}
Ultracold atoms provide a unique toolbox to study many-particle physics under very clean and well-defined conditions. The precise control over their interactions and their trapping potentials allows to study the dynamics of phase transitions as well as the preparation of strongly correlated quantum states \cite{Bloch08}.

While so far the majority of experiments is carried out with ground state atoms, exploiting the unique properties of highly excited states is gradually moving into the focus of experimental and theoretical efforts. Atoms in highly excited states can interact strongly, i.e., the interaction strength can be of the order of several tens of MHz at a distance of several micrometers. The corresponding quantum dynamics takes place on a microsecond timescale and thus is orders of magnitude faster than the atoms' external dynamics. Such scenario is usually referred to as 'frozen gas' \cite{Mourachko98,Anderson98}. A number of experimental groups have studied the excitation dynamics of such system using Rydberg states of alkali metal atoms \cite{Gallagher84} which were excited from an ultracold gas. Here a dramatic reduction of the fraction of excited atoms was observed once the atomic density was too high or the interaction between excited states was too strong \cite{Tong04,Singer04}. This is a manifestation of the so-called 'Rydberg blockade' \cite{Lukin01} effect that is responsible for the collective character \cite{Heidemann07} of Rydberg excitations in dense gases.

Very recently the power of Rydberg states to establish a controlled interaction of single atoms trapped in distant traps has been demonstrated in a series of impressive experiments \cite{Urban08,Gaetan08,Isenhower09,Wilk09}. Strongly supported by these results, highly excited atoms nowadays are believed to have a manifold of applications ranging far beyond traditional atomic physics. Indeed, exploiting the properties of atoms in Rydberg states permits the study of spin systems at criticality \cite{Weimer08}, the quantum simulation of complex spin models \cite{Weimer09}, the investigation of the thermalization of strongly interacting many-particle systems \cite{Olmos09-2} and also the implementation of quantum information protocols \cite{Saffman09-2}.

In a recent work (Ref. \cite{Olmos09-3}) we showed that the unique properties of Rydberg atoms allow the creation of entangled many-particle states on a one-dimensional ring lattice on a short time scale. Finding simple ways for creating entangled many-particle states is of importance, since such states have a number of applications, e.g, they serve as resource for the creation of single-photon light sources \cite{Porras08}, for improving precision quantum measurements \cite{DAriano01} and for measurement based quantum information processing.

In this paper we will go into depth and largely expand on our previous study. We show that excited many-particle states of a laser-driven gas of Rydberg atoms on a ring lattice can be obtained analytically in the limit of strong laser driving. We give a detailed derivation of the system's Hamiltonian in Sec. \ref{sec:System}. The construction of the many-body excitations, their eigenenergies and their correlation properties are analyzed in Sec. \ref{sec:States}. In Sec. \ref{sec:How_address} we discuss thoroughly how these states can be excited in an experiment. We conclude with a summary and outlook in Sec. \ref{sec:Conclusion}.

\section{The system}\label{sec:System}
\begin{figure}
\includegraphics[width=7cm]{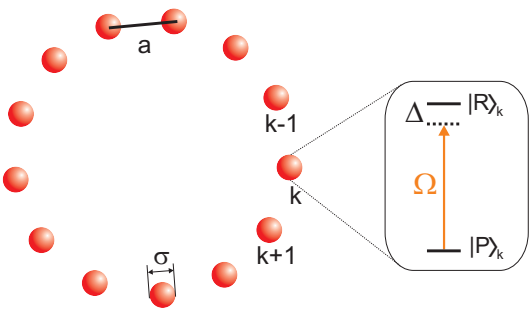}
\caption{Ring lattice with spacing $a$ being much larger than the extension $\sigma$ of the Wannier functions (deep lattice). The internal atomic degrees of freedom at each site are described by the (collective) states $\left|P\right>_k$ and $\left|R\right>_k$, coupled by $\Omega$.}\label{fig:lattice}
\end{figure}

We study a gas of bosonic ground-state atoms confined to a deep large spacing optical or magnetic \cite{Nelson07, whitlock09} ring lattice with periodicity $a\approx\mu m$ (see Fig. \ref{fig:lattice}). The Wannier functions $\Psi_k(\bf{x})$ are localized at the $k$-th site with a width $\sigma \ll a$. We assume the external dynamics of the atoms to be frozen, i.e., no hopping and hence no particle exchange between the lattice sites is present. This is well justified as the internal (electronic) dynamics - in which we are interested here - takes place on a much shorter timescale, of the order of hundred nanoseconds. We consider two electronic levels which are denoted by $\left|g\right>$ and $\left|r\right>$. Here $\left|r\right>$ is a Rydberg $n$s-state which - due to its quantum defect - is well isolated from any other electronic level. It is coupled to the ground state $\left|g\right>$ via a laser with Rabi frequency $\Omega_0$ and detuning $\Delta$. Within the rotating wave approximation, the Hamiltonian describing the coupling of the atoms to the laser field reads (with $\hbar=1$)
\begin{equation}
H_0=\Omega_0\sum_{k=1}^L\left(b_k^{\dagger}r_k+r_k^{\dagger}b_k\right)+\Delta\sum_{k=1}^L n_k,\label{eqn:laser_ham}
\end{equation}
where $b^{\dagger}_k$ and $r^{\dagger}_k$ ($b_k$ and $r_k$) represent the creation (annihilation) of a ground and a Rydberg state, respectively, and $n_k=r^\dagger_k r_k$ stands for the number of atoms in state $\left|r\right>$ at the $k$-th site.
We will consider throughout this paper the case where each lattice site is occupied by the same number of atoms, $N_0$. This is achieved, for
example, if the system is initialized in a Mott-insulator state.

The interaction between the Rydberg atoms is given by the van-der-Waals
potential $C_6/x^{6}$, that is quickly decaying with the
distance $x$ between atoms. Nevertheless, as $C_6$ scales with the eleventh
power of the principal quantum number $n$, the interaction can strongly affect
the excitation dynamics of atoms that are separated by several
micrometers. This strong interaction gives rise to the so-called blockade
effect \cite{Jaksch00,Lukin01}. We consider a scenario in which the
simultaneous excitation of two or more atoms to the Rydberg state on a single
lattice site is blockaded. Thus, on each lattice site $k$, only the two states
\begin{eqnarray*}
\left|P\right>_k&=&\left[\left|g\right>_k\right]_1\otimes\dots\left[\left|g\right>_k\right]_{N_0}\\
\left|R\right>_k&=&\frac{1}{\sqrt{N_0}}{\cal S} \left\{\left[\left|r\right>_k\right]_1\otimes\left[\left|g\right>_k\right]_2\otimes\dots\left[\left|g\right>_k\right]_{N_0}\right\},
\end{eqnarray*}
are accessible, where $\cal S$ is the symmetrization operator.
The effective Rabi frequency for the laser coupling between these so-called (super)atom states (see Fig. \ref{fig:lattice}) is given by $\Omega=\Omega_0\sqrt{N_0}$. Taking all this into account, in Eq. (\ref{eqn:laser_ham}) we can replace
$\Omega_0\, r_k^{\dagger}b_k \rightarrow \Omega\, \sigma^{(k)}_+=(\Omega/2)[\sigma^{(k)}_x + i\sigma^{(k)}_y]$, where $\sigma_x^{(k)}$ and $\sigma_y^{(k)}$ are the Pauli spin matrices.

Since $\sigma \ll a$ (see Fig. \ref{fig:lattice}) we can rewrite the van-der-Waals potential between two (super)atoms in the state $\left|R\right>$ located $d$ sites apart as $\beta_d=C_6/x_d^6$, where $x_d$ is the separation between those sites.
As already pointed out, $\beta_d$ is quickly decaying with the distance. In particular, the next-nearest neighbor interaction is a factor of $64\cos^6{(\pi/L)}$ smaller than the nearest neighbor one ($\beta_2=\beta_1/(2\cos{(\pi/L)})^6$). We will thus only focus on the nearest neighbor interaction which is well-justified for large enough lattices. The interaction Hamiltonian for the entire atomic ensemble, with $\beta\equiv\beta_1$, then reads
\begin{equation*}
H_\mathrm{int}=\beta\sum_{k=1}^Ln_kn_{k+1}
\end{equation*}
with the Rydberg number operator $n_k=[1+\sigma_z^{(k)}]/2$ and the boundary condition $\sigma_j^{(1)}=\sigma_j^{(L+1)}$.

In summary, the complete Hamiltonian that drives the dynamics of our system can be written as
\begin{equation}\label{eqn:working_hamiltonian}
H_\mathrm{spin}=\sum_{k=1}^L\left[\Omega\sigma_x^{(k)}+\Delta n_k+\beta n_k n_{k+1}\right].
\end{equation}
The system can be described as a periodic arrangement of spin-$1/2$ particles, where the two spin states, corresponding to the two internal states of the (super)atoms, $\left|P\right>_k$ and $\left|R\right>_k$, interact via an Ising-type potential. In this picture, the Rabi frequency $\Omega$ and the combination of $\Delta+\beta$ can be effectively interpreted as perpendicular magnetic fields. Hence, the relevant parameters in our system will be: a) the ones related to the laser, i.e., the single-atom Rabi frequency $\Omega_0$ and detuning $\Delta$, which can be time-dependent and b) the interaction between Rydberg atoms represented by $\beta$.

\subsection{Constrained dynamics}\label{sec:Constrained}
Throughout this paper, we consider the regime where the detuning is much smaller than both the collective Rabi frequency (laser driving) and the interaction strength, i.e., $\left|\Delta\right|\ll \Omega,\beta$. As a consequence, the behavior of the system will be determined by the ratio of the latter two parameters. Here we focus on the limit $\Omega\gg\beta$, i.e., the laser coupling is much stronger than the interaction between atoms. In this regime the first term of the Hamiltonian (\ref{eqn:working_hamiltonian}) is the dominant one and it is convenient to make it diagonal by means of a rotation of the basis. This is achieved by the unitary transformation $U=\prod_{k=1}^L \exp{\left(-i\frac{\pi}{4}\sigma_y^{(k)}\right)}$ which brings $\sigma_x\rightarrow\sigma_z$ and $\sigma_z\rightarrow-\sigma_x$. When applied to our Hamiltonian (\ref{eqn:working_hamiltonian}), it yields
\begin{equation}\label{eqn:UHU}
H=U^\dagger H_\mathrm{spin}U=\frac{\beta L}{4}+H_\mathrm{xy}+H_1+H_2,
\end{equation}
with
\begin{eqnarray}
H_\mathrm{xy}&=&\sum_{k=1}^L \left[\Omega\sigma_z^{(k)}+\frac{\beta}{4}\left(\sigma_{+}^{(k)}\sigma_{-}^{(k+1)}+\sigma_{-}^{(k)}\sigma_{+}^{(k+1)}\right)\right]\label{eqn:H_xy}\\
H_1&=&\frac{\Delta}{2}\sum_{k=1}^L\left(1-\sigma_{x}^{(k)}\right)\label{eqn:H_1}\\
H_2&=&\frac{\beta}{4}\sum_{k=1}^L \left[\left(\sigma_{+}^{(k)}\sigma_{+}^{(k+1)}+\sigma_{-}^{(k)}\sigma_{-}^{(k+1)}\right)-2\sigma_x^{(k)}\right]\label{eqn:H_2},
\end{eqnarray}
where $H_\mathrm{xy}$ is the famous $xy$-model of a spin chain with a transverse magnetic field.

\begin{figure}
\includegraphics[width=5cm]{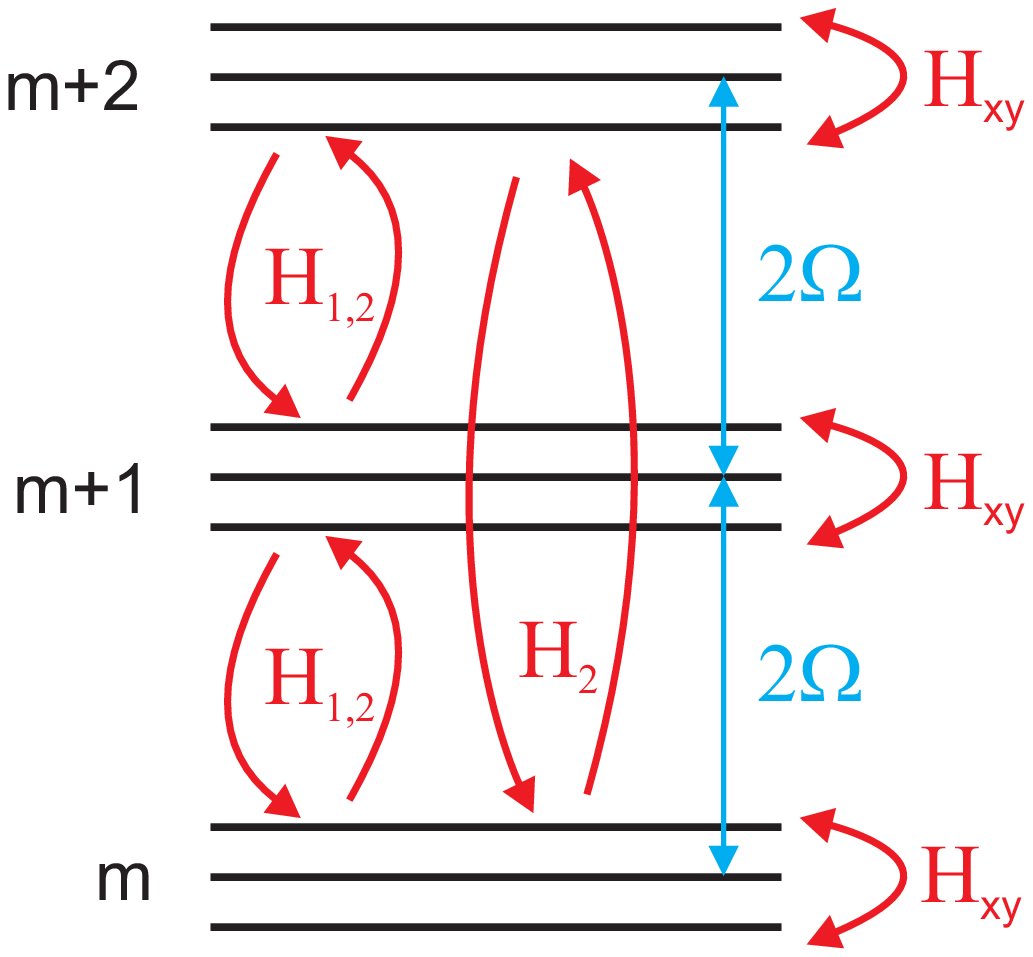}
\caption{Level structure in the regime $\Omega\gg\beta$ and $|\Delta| \ll \Omega$. The spectrum splits into manifolds which can be labeled by the quantum number $m$ of the operator $\sum_k \sigma^{(k)}_z$. For sufficiently large $\Omega$, the coupling between manifolds that is established only by $H_1$ and $H_2$ can be neglected. The (constrained) dynamics inside the $m$-subspaces is then determined by $H_\mathrm{xy}$.}\label{fig:spectrum}
\end{figure}

Let us now analyze the importance of the individual contributions of $H$. As we can see in Fig. \ref{fig:spectrum}, the spectrum of $H$ decays into manifolds of states which are separated by gaps whose width is approximately $2\Omega$. This is caused by the dominant first term of $H_\mathrm{xy}$, i.e., $\Omega\sum_k\sigma_z^{(k)}$. The eigenstates of $\sigma^{(k)}_z$ are - in terms of the (super)atom states - given by
\begin{equation*}
\left|\pm\right>_k=\frac{1}{\sqrt{2}}U^\dagger\left[\left|P\right>_k\pm\left|R\right>_k\right]
\end{equation*}
with $\sigma^{(k)}_z\left|\pm\right>_k=\pm\left|\pm\right>_k$. Thus, each of the manifolds that determine the coarse structure of the spectrum is spanned by a set of product states that have the same number of (super)atoms in the state $\left|+\right>$. In Fig. \ref{fig:spectrum}, these manifolds are denoted by $m$, which is the eigenvalue of the states with respect to the operator $\sum_{k}^L\sigma_z^{(k)}$.

The second term of $H_\mathrm{xy}$ conserves the total number of $\left|+\right>$ (super)atoms. In other words, it couples only states that belong to the same $m$-manifold and that are nearly degenerate. As a consequence, the strength of these intra-manifold couplings due to $H_{\mathrm{xy}}$ is proportional to $\beta$.
Conversely, $H_1$ and $H_2$ couple states that belong to manifolds with different number of (super)atoms in the state $\left|+\right>$.
In particular, $H_1$ and the last term of $H_2$ flip one of the (super)atoms from $\left|+\right>$ to $\left|-\right>$ or viceversa. Thus, the coupled states belong to different manifolds with $\Delta m=\pm1$, energetically separated by $2\Omega$. The two first terms of $H_2$ drive a similar process, flipping always two contiguous (super)atoms in the same state simultaneously, i.e., $\left|++\right>\rightarrow\left|--\right>$ or $\left|--\right>\rightarrow\left|++\right>$. As a result, these terms connect states with eigenvalue $m$ to those with $m\pm2$, which are separated roughly by $4\Omega$. These features are reflected in Fig. \ref{fig:spectrum}.
The transition rates between $m$-manifolds corresponding to $H_1$ and $H_2$ can be estimated by second order perturbation theory to be of the order $\Delta^2/\Omega$ and $\beta^2/\Omega$, respectively. Hence, for sufficiently strong driving $\Omega\gg\beta$, their contribution can be neglected and the system's dynamics is constrained to the $m$-manifolds. As a consequence, the Hamiltonian that drives the intra-manifold dynamics, $H_\mathrm{xy}$, \textit{effectively drives the dynamics of the entire system in this parameter regime}. This Hamiltonian is analytically solvable, and we thus have access to the actual spectrum and eigenstates of the system. The diagonalization of this Hamiltonian relies on the so-called Jordan-Wigner transformation and a Fourier transform that we explain thoroughly in the following paragraph \cite{DePasquale09}.

\subsection{Jordan-Wigner transformation on a ring}
The Pauli matrices in the Hamiltonian (\ref{eqn:H_xy}) obey anti-commutation and  commutation relations when they belong to the same and different sites, respectively. Thus, the algebra is neither bosonic nor fermionic. This difficulty can be overcome by the Jordan-Wigner transformation,
\begin{equation}\label{eqn:J-W}
c_k^{\dagger}=\sigma_+^{(k)}\prod_{j=1}^{k-1}\left(-\sigma_z^{(j)}\right)\quad c_k=\prod_{j=1}^{k-1}\left(-\sigma_z^{(j)}\right)\sigma_-^{(k)},
\end{equation}
which introduces the operators $c_k^{\dagger}$ and $c_k$ that obey the canonical fermionic algebra
\begin{eqnarray*}
\{c^\dagger_i,c_j\}=\delta_{i,j}\qquad\{c^\dagger_i,c^\dagger_j\}=\{c_i,c_j\}=0.
\end{eqnarray*}
After this transformation, the Hamiltonian (\ref{eqn:H_xy}) takes on the form
\begin{eqnarray}\label{eqn:H_JW}\nonumber
H_{\mathrm{xy}}&=&\sum_{k=1}^L \left[2\Omega\left(c_k^\dagger c_k-\frac{1}{2}\right)+\frac{\beta}{4}\left(c_k^\dagger c_{k+1}+c_{k+1}^\dagger c_k\right)\right]\\
&-&\frac{\beta}{4}\left(c_L^\dagger c_1+c_1^\dagger c_L\right)\left(e^{i\pi n_+}+1\right).
\end{eqnarray}
Thus, the Hamiltonian has been transformed into one which describes a chain of spinless fermions with nearest neighbor hopping. The last term of Hamiltonian (\ref{eqn:H_JW}) appears due to the periodic boundary conditions. It depends on the operator $n_+=\sum_{j=1}^{L}c_j^\dagger c_j$ which counts the total number of fermions, which is also equivalent to the number of (super)atoms
in the state $\left|+\right>$.
Thus, depending on the parity of the number of fermions of the state, $H_\mathrm{xy}$ reads
\begin{eqnarray*}
H^\mathrm{(e/o)}_\mathrm{xy}&=&\sum_{k=1}^L 2\Omega\left(c_k^\dagger c_k-\frac{1}{2}\right)+\frac{\beta}{4}\sum_{k=1}^{L-1}\left(c^\dagger_{k}c_{k+1}+c^\dagger_{k+1}c_k\right)\\
&&\mp\frac{\beta}{4}\left(c^\dagger_{L}c_{1}+c^\dagger_{1}c_L\right),
\end{eqnarray*}
for even (e) or odd (o) parity, respectively.

These two cases can be accounted for simultaneously in a convenient way by introducing a matrix representation for the fermionic operators. They are projected onto the subspaces with even and odd eigenvalue of $n_+$ by means of the projectors
$P_\mathrm{e/o}=\left[1\pm e^{i\pi n_+}\right]/2$, with $P_\mathrm{e}+P_\mathrm{o}=\mathbf{1}$. Since the Hamiltonian $H_\mathrm{xy}$ conserves the number of fermions, i.e., $\left[H_\mathrm{xy},e^{i\pi n_+}\right]=0$, it is diagonal in this representation and can be decomposed as
\begin{eqnarray*}
H_{\mathrm{xy}}=\left(\begin{array}{cc}
P_\mathrm{e}H_{\mathrm{xy}}P_\mathrm{e} & P_\mathrm{e}H_{\mathrm{xy}}P_\mathrm{o}\\
P_\mathrm{o}H_{\mathrm{xy}}P_\mathrm{e} & P_\mathrm{o}H_{\mathrm{xy}}P_\mathrm{o}
\end{array}\right)\equiv
\left(\begin{array}{cc}
H_\mathrm{xy}^\mathrm{(e)} & 0\\
0 & H_\mathrm{xy}^\mathrm{(o)}
\end{array}\right).
\end{eqnarray*}
We now introduce new matrix-valued creation and annihilation operators of the form
\begin{eqnarray*}
\gamma^\dagger_k=\left(\begin{array}{cc}
0 & c^\dagger_k\\
c^\dagger_k & 0
\end{array}\right)\qquad
\gamma_k=\left(\begin{array}{cc}
0 & c_k\\
c_k & 0
\end{array}\right),
\end{eqnarray*}
which obey the fermionic algebra provided $c_k$ and $c^\dagger_k$ are fermionic operators. The Hamiltonian can now be conveniently written as
\begin{eqnarray}\nonumber
H_\mathrm{xy}&=&2\Omega\sum_{k=1}^L \left(\gamma^\dagger_k\gamma_k-\frac{1}{2}\right)+\frac{\beta}{4}\sum_{k=1}^{L-1}\left(\gamma^\dagger_k\gamma_{k+1}+\gamma^\dagger_{k+1}\gamma_k\right)\\\label{eqn:H_matrix_gamma}
&&-\frac{\beta}{4}\left(\gamma^\dagger_{L}\gamma_{1}+\gamma^\dagger_{1}\gamma_{L}\right)e^{i\pi n_+},
\end{eqnarray}
with
\begin{eqnarray*}
e^{i\pi n_+}=\left(\begin{array}{cc}
1 & 0 \\
0 & -1
\end{array}\right).
\end{eqnarray*}

The diagonalization of the Hamiltonian (\ref{eqn:H_matrix_gamma}) is achieved
by performing the following Fourier transform
\begin{eqnarray*}
\gamma^\dagger_k=\frac{1}{\sqrt{L}}\sum_{n=1}^L V_{nk}\Lambda^\dagger_n\quad\gamma_k=\frac{1}{\sqrt{L}}\sum_{n=1}^L V^\dagger_{nk}\Lambda_n,
\end{eqnarray*}
with the Fourier coefficients
\begin{eqnarray*}
V_{nk}=\left(\begin{array}{cc}
e^{-i\frac{2\pi}{L}(n-1/2)k} & 0 \\
0 & e^{-i\frac{2\pi}{L}nk}
\end{array}\right).
\end{eqnarray*}
The operators $\Lambda^\dagger_n$ and $\Lambda_n$ are matrix-valued
\begin{eqnarray*}
\Lambda^\dagger_n=\left(\begin{array}{cc}
0 & P_\mathrm{e}\eta^\dagger_{n}\\
P_\mathrm{o}\eta^\dagger_{n} & 0
\end{array}\right)\qquad
\Lambda_n=\left(\begin{array}{cc}
0 & \eta_{n}P_\mathrm{o}\\
\eta_{n}P_\mathrm{e} & 0
\end{array}\right),
\end{eqnarray*}
with $\eta_n^\dagger$ and $\eta_n$ being fermionic creation and annihilation
operators, respectively. Defining the eigenvalue matrix $\epsilon_n$ as
\begin{eqnarray*}
\epsilon_n=2\left(\begin{array}{cc}
\cos{\left[\frac{2\pi}{L}(n-1/2)\right]} & 0\\
0 & \cos{\left[\frac{2\pi}{L}n\right]}
\end{array}\right),
\end{eqnarray*}
the diagonalized Hamiltonian (\ref{eqn:H_matrix_gamma}) reads
\begin{eqnarray}\label{eqn:H_matrix_lambda}
H_\mathrm{xy}=-L\Omega+\sum_{n=1}^L\left(2\Omega+\frac{\beta}{4}\epsilon_n\right)\Lambda^\dagger_n\Lambda_n.
\end{eqnarray}
As we will see in the next section, the introduction of the matrix-valued fermionic operators has the advantage that excited states can be constructed by applying products of $\Lambda^\dagger_n$ to the ground state.
As a consequence, this matrix notation allows us to automatically distinguish between the odd and even fermion cases, which otherwise has to be done manually.

\section{Many-body states}\label{sec:States}
\subsection{Symmetries}
The symmetry properties of our system impose certain selection rules for the excitation
of the many-particle states. In order to understand this, let us start our analysis of the
excited states by studying the symmetries of the Hamiltonian.
Because of the special arrangement of the sites, the Hamiltonian
(\ref{eqn:working_hamiltonian}) and also (\ref{eqn:H_matrix_lambda}) are
invariant under cyclic shifts and reversal of the lattice sites. This can be formally
seen by representing these two symmetries through the operators
$\cal{X}$ and $\cal{R}$, respectively.
Their action on the spin ladder operators are ${\cal
  X}^\dagger\sigma_\pm^{(k)}{\cal X}=\sigma_\pm^{(k+1)}$ and ${\cal
  R}^\dagger\sigma_\pm^{(k)}{\cal R}=\sigma_\pm^{(L-k+1)}$, from where follows
that $\left[H_\mathrm{spin},{\cal X}\right]=\left[H_\mathrm{spin},{\cal R}\right]=0$, i.e.,
both of them correspond to conserved quantities.
Thus, if the system is initialized in an eigenstate with respect to ${\cal X}$ and
${\cal R}$, the time evolution will not take place in the entire Hilbert space,
but merely in the subspace spanned by the states with the same quantum number
with respect to $\cal{X}$ and $\cal{R}$.

This observation is highly relevant for our system. In practise, the natural
initial situation will be that in which all atoms are in the ground state, i.e.,
$\left|0\right>=\prod_{k=1}^L\left|P\right>_k$. This state has the
above-mentioned properties, i.e., it is invariant under cyclic shifts and the
reversal of the sites: ${\cal X}^\dagger\left|0\right>=\left|0\right>$ and
${\cal R}^\dagger\left|0\right>=\left|0\right>$. We will refer to such a state
that has eigenvalue $1$ with respect to ${\cal X}$ and ${\cal R}$ as being
fully-symmetric.
Hence, only the states from this fully-symmetric subspace can
be actually accessed in the course of the system's time evolution under
Hamiltonian (\ref{eqn:working_hamiltonian}).
In the following we will thus
focus on constructing excited states that belong to this subset.

\subsection{Fully-symmetric states}
The ground state of Hamiltonian (\ref{eqn:H_matrix_lambda}) is given by
\begin{eqnarray*}
\left|G\right>=\prod_{k=1}^L\left|-\right>_k
\end{eqnarray*}
and it is fully-symmetric. Excited states that contain $N$ fermions are in general formed by successive application of the creation operator, i.e., $\left|N_{pq\dots}\right>=\Lambda_p^\dagger\Lambda_q^\dagger\dots\left|G\right>$. However, not all combinations will give rise to states that belong to the fully-symmetric subset.

Let us start considering the possible cases of a single-fermion excitation. For a fully-symmetric state we require $O^\dagger\left|1_p\right>=O^\dagger\Lambda_p^\dagger\left|G\right>=O^\dagger\Lambda_p^\dagger O\left|G\right>=\Lambda_p^\dagger\left|G\right>=\left|1_p\right>$, i.e.,
\begin{eqnarray*}
O^\dagger\Lambda_p^\dagger O\left|G\right>=\Lambda_p^\dagger\left|G\right>,
\end{eqnarray*}
with $O$ being a placeholder for $\cal{X}$ and $\cal{R}$. After some algebra one finds that
\begin{eqnarray*}
{\cal R}^\dagger \eta_p^\dagger{\cal R}&=&e^{i\frac{2\pi}{L}p}\eta_{L-p}^\dagger e^{i\pi n_+}\\
{\cal X}^\dagger \eta_p^\dagger{\cal X}&=&e^{-i\frac{2\pi}{L}p}\eta_{p}^\dagger e^{i\pi c_1^\dagger c_1}+\frac{1}{\sqrt{L}}c_1^\dagger\left(e^{i\pi n_+}-1\right).
\end{eqnarray*}
Since $e^{i\pi n_+}\left|G\right>=\left|G\right>$ and $e^{i\pi c_ 1^\dagger c_1}\left|G\right>=\left|G\right>$, only the single excitation with $p=L$ is symmetric under cyclic shifts and reversal. Hence, the only one-fermion state that can be reached by the time-evolution reads
\begin{eqnarray*}
\left|1\right>=\Lambda_L^\dagger\left|G\right>.
\end{eqnarray*}
To have a better physical understanding of this state, it is convenient to write it in terms of the atomic operators,
\begin{eqnarray*}
\left|1\right>=\frac{1}{\sqrt{L}}\sum_{k=1}^L\sigma_+^{(k)}\left|G\right>.
\end{eqnarray*}
Thus, $\left|1\right>$ is a spin wave or, in other words, a superatom that extends over the entire lattice. These states are of interest since they can be used as a resource for single photon generation.

For the two-fermion states, we follow the same procedure and demand
\begin{eqnarray*}
O^\dagger\Lambda_p^\dagger \Lambda_q^\dagger O\left|G\right>=\Lambda_p^\dagger \Lambda_q^\dagger \left|G\right>.
\end{eqnarray*}
One finds that
\begin{eqnarray*}
{\cal R}^\dagger \eta_p^\dagger \eta_q^\dagger{\cal R}&=&e^{i\frac{2\pi}{L}(p+q-1)}\eta_{L-q+1}^\dagger \eta_{L-p+1}^\dagger\\
{\cal X}^\dagger \eta_p^\dagger \eta_q^\dagger{\cal X}&=&e^{-i\frac{2\pi}{L}(p+q-1)}\eta_{p}^\dagger \eta_{q}^\dagger \\
&&+\frac{e^{i\frac{\pi}{L}}}{\sqrt{L}}\left[e^{-i\frac{2\pi}{L}p}\eta_{p}^\dagger-e^{-i\frac{2\pi}{L}q}\eta_{q}^\dagger\right]c_1^\dagger\left(e^{i\pi n_+}-1\right).
\end{eqnarray*}
From this, one sees that the condition $p+q-1=L$ has to be accomplished. As a result, the fully-symmetric states are
\begin{eqnarray*}
\left|2_p\right>=\Lambda^\dagger_p\Lambda^\dagger_{L-p+1}\left|G\right>,
\end{eqnarray*}
with $p=1\dots\lfloor L/2\rfloor$. These are entangled states formed by superpositions of two-atom excitations in the ring with opposite momentum. This is more clearly seen by writing everything in terms of the Pauli matrices
\begin{eqnarray*}
\left|2_p\right>=\frac{2}{iL}\sum_{k>k'}\sin{\left(\frac{2\pi}{L}(p-1/2)(k-k')\right)}\sigma_+^{(k)}\sigma_+^{(k')}\left|G\right>.
\end{eqnarray*}
These states are potentially interesting for the production of photon pairs.
How they can be actually accessed will be discussed in Sec. \ref{sec:How_address}.

Finally, let us illustrate how the three-fermion excitations are formed. We have
\begin{eqnarray*}
R^\dagger \eta^\dagger_p\eta^\dagger_q\eta^\dagger_r R&=&e^{i\frac{2\pi}{L}(p+q+r)}\eta^\dagger_{L-p}e^{i\pi n_+}\eta^\dagger_{L-q}e^{i\pi n_+}\eta^\dagger_{L-r}e^{i\pi n_+}\\
&=&-e^{i\frac{2\pi}{L}(p+q+r)}\eta^\dagger_{L-p}\eta^\dagger_{L-q}\eta^\dagger_{L-r}e^{i\pi n_+}.
\end{eqnarray*}
and thus fully symmetric three-fermion states are of the form
\begin{eqnarray}
\left|3_{pqr}\right>=\frac{1}{\sqrt{2}}\left(\Lambda^\dagger_{p}\Lambda^\dagger_{q}\Lambda^\dagger_{r}-\Lambda^\dagger_{L-p}\Lambda^\dagger_{L-q}\Lambda^\dagger_{L-r}\right)\left|G\right>,
\end{eqnarray}
with $p+q+r=L,2L$. Writing these eigenexcitations back in terms of the spin operators yields
\begin{eqnarray*}
\left|3_{pqr}\right>&=&-\frac{\sqrt{2}i}{L^{3/2}}\sum_{k>k'>k''}\sum_{\begin{subarray}{c}\mathrm{perm}\\(pqr)\end{subarray}}\varepsilon_{pqr}\\
&&\times\sin{\left[\frac{2\pi}{L}(kp+k'q+k''r)\right]}\sigma_+^{(k)}\sigma_+^{(k')}\sigma_+^{(k'')}\left|G\right>,
\end{eqnarray*}
where $\varepsilon_{pqr}$ is the Levi-Civita symbol.
In a similar way, states with higher number of fermions are obtained.

\subsection{Energy spectrum}\label{sec:energy}
Now that we have analyzed the eigenstates of the system we will focus on the corresponding eigenenergies. In the course of this investigation we will also perform a comparison of the analytic results to the ones obtained from a numerical diagonalization of the Hamiltonian (\ref{eqn:working_hamiltonian}). This will allow us to assess the accuracy of our analytical approach.

\begin{figure}
\includegraphics[width=9cm]{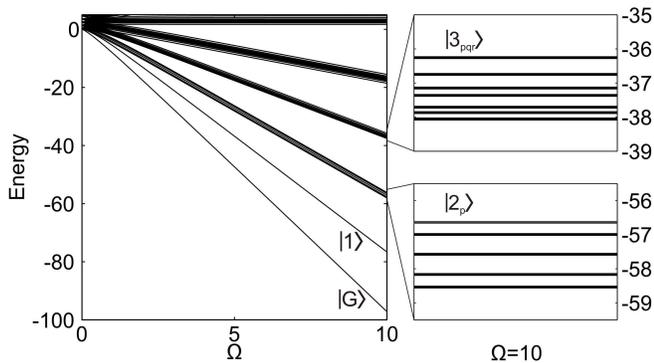}
\caption{Spectrum of Hamiltonian $H$ (\ref{eqn:UHU}) for a lattice of $L=10$ sites versus the laser driving $\Omega$ in units of $\beta$. In the right insets, the energies of the two- and three-fermion states are shown for $\Omega=10$. Five two-fermion and eight three-fermion eigenenergies arise as it is analytically predicted for this lattice size.}\label{fig:spectrum_real}
\end{figure}

Let us begin with the ground state energy. From Eq. (\ref{eqn:H_matrix_lambda}) we can read off the value
\begin{eqnarray}\label{eqn:E_G}
E_{G}=-L\left(\Omega-\frac{\beta}{4}\right),
\end{eqnarray}
where we have included the general energy-offset $\beta L/4$ (see Eq. (\ref{eqn:UHU})).
For $\Delta=0$, $\Omega=10$, $\beta=1$ and $L=10$, the result is
$E_G=-97.5$. This is to be compared with the numerical
value of $-97.63$ which is obtained by diagonalizing the Hamiltonian
(\ref{eqn:working_hamiltonian}).
We find both results to be in good agreement. For the first excited state we obtain
\begin{eqnarray*}
E_1=E_G+2\Omega+\frac{\beta}{2}.
\end{eqnarray*}
Using the same set of parameters, the energy of the single-fermion state is $E_1=-77.0$, which is very close to the numerically exact value $-77.11$. The energies of higher eigenexcitations are given by
\begin{eqnarray*}
E_{2p}=E_G+4\Omega+\beta\cos{\left[\frac{2\pi}{L}(p-1/2)\right]},
\end{eqnarray*}
with $p=1\dots\lfloor L/2\rfloor$, for the two-fermion case and
\begin{eqnarray*}
E_{3pqr}&=&E_G+6\Omega+\frac{\beta}{2}\left[\cos{\left(\frac{2\pi}{L}p\right)}\right.\\
&&\left.+\cos{\left(\frac{2\pi}{L}q\right)}+\cos{\left(\frac{2\pi}{L}r\right)}\right],
\end{eqnarray*}
with $p+q+r=L,2L$, for the three-fermion one.
For $L=10$, we obtain five and eight different eigenenergies for the two- and three-fermion states, respectively (see insets in Fig. \ref{fig:spectrum_real}). In the Tables \ref{tab:2n} and \ref{tab:3lmn} we perform a comparison between the analytical and the numerical results. A difference of less than a $0.2\%$ is observed in all cases.
\begin{table}
\begin{tabular}{|c|c|c|}
\hline
$p$ & $E_{2p}$ & Numerical \\
\hline
1 & -56.55 & -56.64 \\
2 & -56.91 & -56.99 \\
3 & -57.50 & -57.58 \\
4 & -58.09 & -58.17 \\
5 & -58.45 & -58.54 \\
\hline
\end{tabular}
\caption{Energies of the five two-fermion states $\left|2_p\right>$ for $L=10$, $\Delta=0$, $\Omega=10$ and $\beta=1$ and comparison with the numerically exact values.}\label{tab:2n}
\end{table}
\begin{table}
\begin{tabular}{|c|c|c|c|c|}
\hline
$p$ & $q$ & $r$ & $E_{3pqr}$ & Numerical \\
\hline
1 & 9 & 10& -36.19 & -36.26 \\
2 & 8 & 10& -36.69 & -36.75 \\
1 & 2 & 7 & -37.10 & -37.15 \\
3 & 7 & 10& -37.31 & -37.36 \\
1 & 3 & 6 & -37.65 & -37.71 \\
4 & 6 & 10& -37.81 & -37.87 \\
1 & 4 & 5 & -38.00 & -38.05 \\
2 & 3 & 5 & -38.00 & -38.06 \\
\hline
\end{tabular}
\caption{Energies of the eight three-fermion states $\left|3_{pqr}\right>$ for $L=10$, $\Delta=0$, $\Omega=10$ and $\beta=1$ and comparison with the numerically exact values.}\label{tab:3lmn}
\end{table}

The discrepancies between analytical and numerical values are mainly caused by second order energy shifts due to $H_1$ and $H_2$ (Eqs. (\ref{eqn:H_1}) and (\ref{eqn:H_2})). These contributions vanish only in the limits $\beta/\Omega\rightarrow 0$ and $\Delta/\Omega\rightarrow 0$. Here, we will calculate them for a finite ratio. There is a constant term in $H_1$ which is proportional to $\Delta$ that gives rise to a global energy shift $E^{(1)}=L\Delta/2$. Being aware of this shift facilitates the comparison between the numerically exact and the approximate analytical eigenvalues for   $\Delta\neq 0$.

Let us focus first on the ground state. $H_1$ and $H_2$ only
couple states whose number of fermions differ by one or two
(Fig. \ref{fig:spectrum}). As a consequence, only the states $\left|1\right>$ and
$\left|2_p\right>$ contribute to the second order correction of the energy of
the ground state. It yields
\begin{eqnarray*}
E_G^{(2)}&=&-\frac{L\left|\Delta+\beta\right|^2}{8\Omega+2\beta}-\frac{\beta^2}{4}\left(1+\frac{2}{L}\right)^2\\
&&\times\sum_{p=1}^{\lfloor L/2\rfloor}\frac{\sin^2{\left[\frac{2\pi}{L}(p-1/2)\right]}}{4\Omega+\beta\cos{\left[\frac{2\pi}{L}(p-1/2)\right]}}.
\end{eqnarray*}

Analogously, we calculate the energy shift of the first excited state, $\left|1\right>$, due to $H_1$ and $H_2$. In this case, we have to compute the effect of the states $\left|G\right>$, $\left|2_p\right>$ and $\left|3_{pqr}\right>$. The resulting energy correction is given by
\begin{eqnarray*}
E_1^{(2)}&=&\frac{L\left|\Delta+\beta\right|^2}{8\Omega+2\beta}-\frac{\left|\Delta+\beta\right|^2}{L}\\
&&\times\sum_{p=1}^{\lfloor L/2\rfloor}\frac{\cot^2{\left[\frac{\pi}{L}(p-1/2)\right]}}{2\Omega+\frac{\beta}{2}\left(2 \cos{\left[\frac{2\pi}{L}(p-1/2)\right]}-1\right)}.
\end{eqnarray*}

For the parameters $\Delta=0$, $\Omega=10$, $\beta=1$ and $L=10$, these shifts yield $E_G^{(2)}=-0.14$ and $E_1^{(2)}=-0.10$. The corrected energies of the ground and the single-fermion state are now $E_G=E_G^{(0)}+E_G^{(1)}+E_G^{(2)}=-97.64$ and $E_1=E_1^{(0)}+E_1^{(1)}+E_1^{(2)}=-77.10$, much closer to the numerically exact ones of $-97.63$ and $-77.11$, respectively. We will later see that these energy corrections can be useful for the selective excitation of many-particle states in the lattice.

\subsection{Correlation functions}
In this subsection we are going to study the density-density correlation function of the many-particle states. This quantity measures the conditional probability of finding two simultaneously excited atoms at a distance $x$ from each other normalized to the probability of uncorrelated excitation. It is defined - for a fully-symmetric state $\left|\Psi\right>$ - as
\begin{equation*}
g_2(x,\Psi)=\frac{\left<n_1n_{1+x}\right>_\Psi}{\left<n_1\right>_\Psi^2}-1,
\end{equation*}
where we have used $\left<n_a\right>_\Psi=\left<n_b\right>_\Psi$ for all sites.
The correlation function will give $g_2(x,\Psi)=0$ when two sites separated by a distance $x$ are completely uncorrelated, and
$g_2(x,\Psi)>0 \, (<0)$ for correlation (anticorrelation) between the sites.

\begin{figure}
\includegraphics[width=9cm]{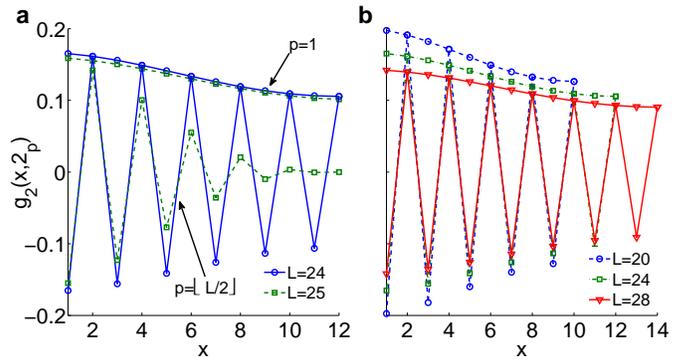}
\caption{Density-density correlation functions of the $\left|2_p\right>$ states. \textbf{a}: For $p=1$ and $p=\lfloor L/2\rfloor$, the correlations show completely different behavior, i.e., smoothly decreasing and strongly oscillating, respectively. These oscillations are much more pronounced for the even value of $L=24$ than for the odd, $L=25$. \textbf{b}: The magnitude of the correlations decreases as the number of sites $L$ is enhanced, as can be seen for $L=20,24,28$.}\label{fig:correlations}
\end{figure}

In particular, for the case $\left|\Psi\right>=\left|2_p\right>$, $g_2(x,2_p)$ can be analytically calculated.
In terms of the expectation values of the spin operators, the correlation function reads
$g_2(x,2_p)=\left<\sigma_+^{(1)}\sigma_-^{(1+x)}+\sigma_-^{(1)}\sigma_+^{(1+x)}\right>_{2_p}$.
For  $x=0$ we have $g_2(0,2_p)=1$ and for $x>0$ the calculation yields
\begin{eqnarray*}
g_2(x,2_p)&=&\frac{4}{L^2}\left[(L-2x)\cos{\left[\frac{2\pi}{L}(p-1/2)x\right]}\right.\\ &&\left.+2\sin{\left[\frac{2\pi}{L}(p-1/2)x\right]}\cot{\left[\frac{2\pi}{L}(p-1/2)\right]}\right].
\end{eqnarray*}
By inspecting this expression for the allowed values
$p=1,\dots\lfloor L/2\rfloor$, some general statements can be made:

a) Independently of the total number of sites $L$, there are always two 'extremal' cases (see Fig. \ref{fig:correlations}a) which correspond to $p=1$ and $p=\lfloor L/2\rfloor$: for $p=1$, the correlation function shows a positive maximum at $x=1$, i.e., nearest neighbor, and then decreases monotonically and smoothly with the distance, staying always positive; for $p=\lfloor L/2\rfloor$, the nearest neighbor is pronouncedly anticorrelated, the next-nearest neighbor is correlated and this pattern of correlation-anticorrelation persists with increasing distance.

b) For  $p=\lfloor L/2\rfloor$, the oscillations of  $g_2(x,2_p)$ are more pronounced for $L$ than for odd $L$, see Fig. \ref{fig:correlations}a.
The ratio of the amplitudes of the correlations for $x=1$ and $x=\lfloor L/2\rfloor$ is,
\begin{displaymath}
\frac{g_2(1,2_{\frac{L}{2}})}{g_2(\frac{L}{2},2_{\frac{L}{2}})}\sim 1
\qquad\frac{g_2(1,2_{\frac{L-1}{2}})}{g_2(\frac{L-1}{2},2_{\frac{L-1}{2}})}\sim L^{3},
\end{displaymath}
in the even and odd cases, respectively.
Also, for an even number of sites, the correlation functions of the two extreme cases accomplish $g_2(x,2_{\frac{L}{2}})=(-1)^xg_2(x,2_1)$, i.e.,
the envelope of the oscillating function  $g_2(x,2_{\frac{L}{2}})$ is given by the smoothly decreasing $g_2(x,2_1)$.

c) For a fixed value of $p$, the amplitude of the correlations decreases with increasing number of sites as $1/L$, as can be seen in Fig. \ref{fig:correlations}b.

Numerically, we have observed agreement to the analytical results
shown in Fig. \ref{fig:correlations}. As expected, this agreement improves with a decreasing ratio $\beta/\Omega$.
The correlations could be directly monitored experimentally provided that a
site-resolved detection of atoms in the $\left|+\right>$-state is possible.
The next section will deal with the open question of how these correlated
states can be experimentally accessed.

\section{Excitation of many particle states}\label{sec:How_address}
Our aim is to selectively excite correlated many-body states by a temporal variation of the laser parameters. Initially the atoms shall be in the product state $\left|0\right>=\prod_{k=1}^L \left|P\right>_k$ and the laser shall be turned off, i.e., $\Omega_0(0)=0$ and $\Delta(0)=\Delta_0$. Starting from these initial conditions, the goal is to vary $\Omega_0(t)$ and $\Delta(t)$ such that at the end of the sequence, i.e., at $t=t_\mathrm{final}$, the detuning is zero and the laser driving is much larger than the interaction ($\Delta(t_\mathrm{final})=0$ and $\Omega(t_\mathrm{final})/\beta\gg 1$). This final situation corresponds to the right-hand side of the spectrum
presented in Fig. \ref{fig:spectrum_real}.

Once a desired many-particle state has been populated, and due to the limited lifetime of the highly excited levels which is in the order of several $\mu s$ (e.g., 66 $\mu s$ for Rb in the 60s state), we want to map it to an stable configuration. To do so, we first turn off the laser ($\Omega=0$) and then switch on a second one whose action can be described by the Hamiltonian
\begin{equation}\label{eqn:H_map}
H_\mathrm{map}=\Omega_s\sum_{k=1}^L\left(s_k^{\dagger}r_k+r_k^{\dagger}s_k\right)+\beta\sum_{k=1}^Ln_kn_{k+1}.
\end{equation}
In this expression, $s^\dagger_k$ and $s_k$ stand for the creation and annihilation operators
of an single-atom stable storage state $\left|s\right>$ on site $k$, respectively.
In the limit where the interaction is much smaller than the Rabi frequency of this transition, i.e., $\Omega_s\gg\beta$, we can neglect the second term of this Hamiltonian. Thus, performing a global $\pi$-pulse to the considered many-particle state means to perform the mapping $\left|r\right>\rightarrow \left|s\right>$, such that a stable configuration is achieved.

Hence, the difficulty lies in finding a 'trajectory' or sequence $(\Omega_0(t),\Delta(t))$ for which at $t=t_\mathrm{final}$ only a single many-particle state is occupied. We propose two different methods in the following.

\subsection{Direct trajectory}
\begin{figure}
\includegraphics[width=9cm]{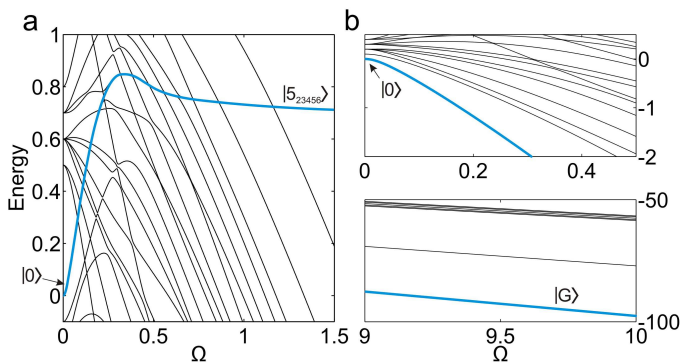}
\caption{Possible trajectories ($\Delta(t),\Omega_0(t)$) through the spectrum
of $H$ with $\Delta_0\ne0$
  (units of $\beta$). \textbf{a}: When $\Delta_0<0$, the ground state at $\Omega=0$ does not coincide with the initial state, $\left|0\right>$, and the energy of the initial state goes through a number of avoided crossings. A possible path through them to reach the state $\left|5_{23456}\right>$ is shown. \textbf{b}: If $\Delta_0>0$, the initial state $\left|0\right>$ is adiabatically connected to the ground state $\left|G\right>$.}\label{fig:address}
\end{figure}

In certain cases, one can guess a trajectory $(\Delta(t),\Omega_0(t))$ like
the ones shown in Fig. \ref{fig:address} that eventually connects
$\left|0\right>$ with a desired eigenstate of $H_\mathrm{xy}$ \cite{Pohl09,Schachenmayer09}, but this is not always possible. The general appearance of the laser sequence strongly depends on the sign of the initial detuning $\Delta_0$. In Fig. \ref{fig:address} the two possible scenarios (taking $\Delta_0\neq 0$) are depicted.
For $\Delta_0<0$, the initial state is not the ground state of the system when the laser is
turned off ($\Omega=0$). As a consequence, this initial state suffers several avoided crossings with other levels
when $\Omega$ is increased. Thus, it is not easy to find a path through the spectrum that connects it to a single desired eigenstate of $H_\mathrm{xy}$, as the one shown in Fig. \ref{fig:address}a.
A more general framework for finding a proper trajectory is provided by Optimal Control theory \cite{Peirce88}. Here, the desired fidelity with which the final state is achieved can be set and certain constraints on the trajectory can be imposed. This method is successfully applied to quantum information processing \cite{Calarco04}, molecular state preparation \cite{Somloi93} and optimization of number squeezing of an atomic gas confined to a double well potential \cite{Grond09}. The case of $\Delta_0>0$ will be treated in the next subsection.

\subsection{Excitation from the ground state}
We present in this work a different route to populate single many-particle states. This is accomplished in two steps: First, one has to prepare the ground state $\left|G\right>$ of Hamiltonian (\ref{eqn:H_xy}) in the limit $\Omega\gg\beta$; once the ground state is populated, the single-fermion and two-fermion many-particle states can be accessed by means of an oscillating detuning, that gives rise to a time-dependent $H_1$.

\begin{figure}
\includegraphics[width=9cm]{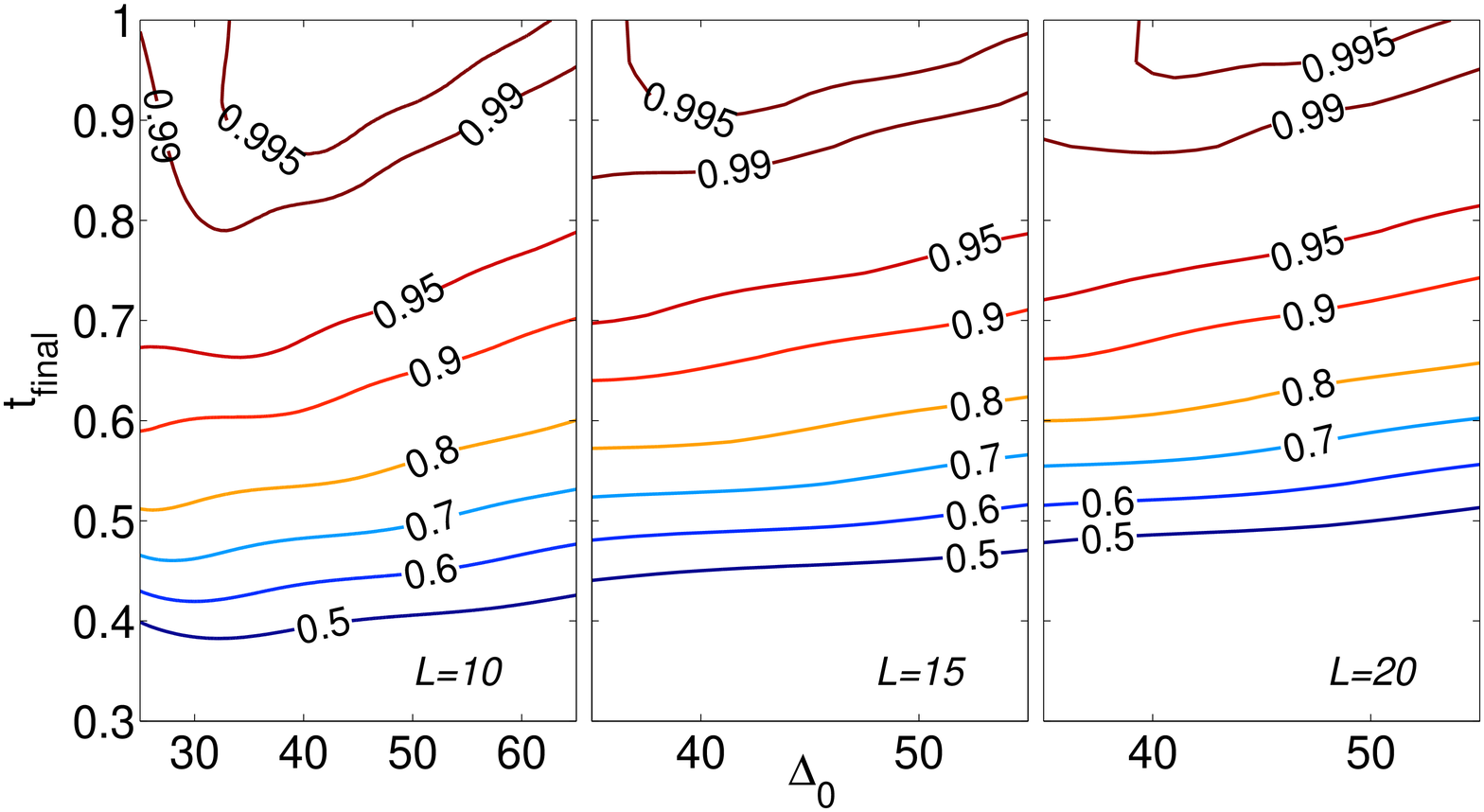}
\caption{Fidelity $\left|\left<\Phi(t_\mathrm{final})\vert G\right>\right|^2$ when populating the ground state of $H_\mathrm{xy}$ from the initial state via variation of the parameters of the laser $\Omega(t)$ and $\Delta(t)$ in the form given by ($\ref{eqn:omega}$) and (\ref{eqn:detuning}), respectively. Several initial values of the detuning and time intervals, as well as different lattice sizes, are considered. For a fixed value of $\Delta_0$ (units of $\beta$), better fidelities are obtained for larger time intervals (units of $1/\beta$). For a fixed time interval, there is an optimal value of $\Delta_0$ for each size of the lattice, around $\Delta_0\approx45$.}\label{fig:0toGS}
\end{figure}

\textit{Step 1:}
Let us start by explaining how to vary the laser parameters to prepare the
ground state $\left|G\right>$.
In particular, when setting $\Delta_0>0$, the ground state of the
system at $\Omega=0$ coincides with the initial state $\left|0\right>$.
With increasing $\Omega$, it is adiabatically connected to the
ground state $\left|G\right>$ of $H_{\mathrm{xy}}$ (see
Fig. \ref{fig:address}b).
The problem that we can encounter here is that non-adiabatic transitions to other energy levels occur when increasing $\Omega$, so that we do not populate only $\left|G\right>$ but also other states.
To avoid this, we choose a large enough value of $\Delta_0$ when the laser is still turned off ($\Omega=0$). This increases the energy gap between  $\left|0\right>$  and other energy levels,  and, as a consequence,  suppresses non-adiabatic transitions.
This initial detuning can be decreased as $\Omega$ increases so that in the desired regime, i.e., $\Omega_\mathrm{final}\equiv\Omega(t_\mathrm{final})\gg\beta$, it is set to zero. As an example, we propose the following shapes of $\Omega(t)$ and $\Delta(t)$
\begin{eqnarray}
\Omega(t)&=&\Omega_\mathrm{final}\sin^2{\left(\frac{\pi t}{2t_\mathrm{final}}\right)}\label{eqn:omega}\\
\Delta(t)&=&\Delta_0\left[1-\sin^2{\left(\frac{\pi t}{2t_\mathrm{final}}\right)}\right]\label{eqn:detuning}.
\end{eqnarray}
The obtained fidelity $\left|\left<\Phi(t_\mathrm{final})\vert G\right>\right|^2$ for different values of the initial detuning $\Delta_0$ and time intervals $t_\mathrm{final}$ is given in Fig. \ref{fig:0toGS}, where $\Phi(t_\mathrm{final})$ stands for the wavefunction of the final state.
It is actually possible to populate the desired state with high fidelity, e.g., over $99\%$ is achieved for all considered lattice sizes with $\Delta_0=45\beta$ and $t_\mathrm{final}=0.9 \beta ^{-1}$.
We find that: i) the fidelity depends only weakly on the lattice size
although the dimension of the Hilbert space grows exponentially with $L$,
and ii) as expected, for a fixed value of the initial detuning, the fidelity increases with the increasing length of the time interval.
Note that the timescale of this whole process is limited by the lifetime of the Rydberg state.

If there is only one atom per site, and based on the fact that  $\left|G\right>=\prod_{k=1}^L\left|-\right>_k$ is a product state, an alternative procedure to this adiabatic passage can be envisaged.
Starting from the vacuum $\left|0\right>$ (also a product state with every atom in $\left|g\right>$), we perform a global $\pi/2$-pulse to the single-atom transition $\left|g\right>\rightarrow\left|s\right>$. As a result, we obtain a product state where every atom is in a superposition $\left[\left|g\right>+i\left|s\right>\right]/\sqrt{2}$.
In a second step, the $\pi$-pulse with the mapping laser described by the Hamiltonian (\ref{eqn:H_map}) and with $\Omega_s\gg\beta$, transfers every atom to the state $\left[\left|g\right>-\left|r\right>\right]/\sqrt{2}$, i.e., we have prepared the ground state $\left|G\right>$.
It is worth remarking that this method eliminates the lifetime limitation in this first stage.

\textit{Step 2:} Let us show now how to address the single-fermion and
two-fermion states from this ground state $\left|G\right>$. As we explained in
section \ref{sec:Constrained}, the Hamiltonian $H_1$, associated with the
detuning, drives transitions between neighboring manifolds, i.e. $\Delta
m=\pm1$, (see Fig. \ref{fig:spectrum}).
We exploit this fact and introduce an oscillating detuning of the form
$\Delta(t)=\Delta_\mathrm{osc}\cos{\left(\omega t\right)}$. If we tune
$\omega$ to coincide with the gap between two given states, this detuning acts
effectively as a laser that couples them resonantly with a Rabi frequency that
is proportional to $\Delta_\mathrm{osc}$.

\begin{figure}
\includegraphics[width=8.5cm]{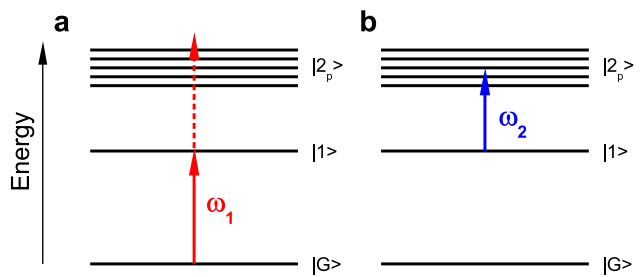}
\caption{Sketch of the excitation of the single-fermion and two-fermion states by means of an oscillating (radiofrequency) detuning using a not too large value of $\Omega$. \textbf{a:} In a first step, the population is transferred by a $\pi$-pulse to the single-fermion state by tuning the frequency of the detuning on resonance with the gap $\omega=\omega_1$. \textbf{b:} A second $\pi$-pulse with $\omega$ tuned to match $\omega_2$ addresses the corresponding $\left|2_p\right>$ state, bearing in mind that $\beta\gg\Delta_\mathrm{osc}$ in this step.}\label{fig:excitation}
\end{figure}

Using this oscillating detuning, we want to transfer the population from the
ground to the first excited state (Fig. \ref{fig:excitation}a). To do so,
$\omega$ is tuned to be on resonance with the corresponding energy gap, i.e.,
$\omega=\omega_1=E_1-E_G$, and by a $\pi$-pulse we populate $\left|1\right>$.
One has to take into account that in the limit of $\Omega\gg\beta$ the energy
gap between any two neighboring manifolds is equal, i.e., also higher lying excitations are populated. To avoid this effect and address only the  $\left|1\right>$ state, we can choose a not too large value of $\Omega$. In this regime, the second order level shifts caused by $H_1$ and $H_2$, that are roughly given by $\Delta^2/\Omega$ and $\beta^2/\Omega$, respectively (see section \ref{sec:energy}), become increasingly important. In particular, as it is sketched in Fig. \ref{fig:excitation}a, the gap between
$\left| 1\right>$ and any of the  $\left|2_p\right>$ levels
becomes more and more different from $\omega_1$ and, as a consequence, the unwanted transitions fall out of resonance.
Analogously, the same procedure could be used to address the two-fermion many-particle states (see Fig. \ref{fig:excitation}b).
The first $\pi$-pulse resonant with the $\left|G\right>\rightarrow\left|1\right>$ transition, is followed by another $\pi$-pulse with $\omega$ tuned to coincide with the energy gap of the specific $\left|1\right>\rightarrow\left|2_p\right>$ transition, $\omega=\omega_2=E_{2p}-E_1$.
The separation between neighboring $\left|2_p\right>$ states is of the order of $\beta$ and the Rabi frequency of the transition is proportional to $\Delta_\mathrm{osc}$. As a consequence, to populate only a single level of the two-fermion manifold, the parameters have to accomplish that $\beta\gg\Delta_\mathrm{osc}$ and, at the same time, $\Delta_\mathrm{osc}$ has to be large enough in order to perform the transfer at a time interval that is much shorter than the lifetime of the Rydberg state.

\section{Conclusions and outlook}\label{sec:Conclusion}
In this work we have studied the collective excitation of a laser-driven
Rydberg gas confined to a ring lattice. We have focused on the regime in which
the interaction between the highly excited states is much weaker than the
laser field. We found that the corresponding system can be described as a chain of spinless fermions whose dynamics is driven by the $xy$-model. This Hamiltonian can be analytically solved and, by exploiting the symmetries of the system, we were able to completely characterize the many-particle states arising.
In particular, we have shown that the first excited state of the Hamiltonian corresponds to a spin wave or to an excitation which is completely delocalized all over the lattice. The two-fermion states could be expressed as a superposition of excitation pairs and an investigation of their density-density correlation function has been performed. We have demonstrated that the qualitative behavior of these correlations differs substantially from one state to another of the same two-fermion manifold, going from a smoothly decaying function to a pronounced correlation-anticorrelation pattern.
The analytical eigenenergies of the $xy$-Hamiltonian were compared to the
numerical exact ones of the complete Hamiltonian, and excellent agreement between both results has been found.
Finally, we have investigated several paths for the selective excitation of the many-particle states. One of them relies on the variation of the laser parameters with time, finding trajectories from the initial to a given final many-body state. The other possibility we have presented makes use of an oscillating detuning which allows to access excitations starting from the ground state of the Hamiltonian. In each step, a $\pi$-pulse is performed with the frequency of the oscillation matching the energy gap between the involved states.

Throughout this work we have considered an homogeneous occupation of the sites of the ring lattice. The situation of having a randomly fluctuating number of atoms per site would effectively lead to a disorder potential for the fermions, as outlined in Ref. \cite{Olmos09-3}. This would imply as well a change in the symmetry properties of the system, so that more states become accessible by a time-evolution (e.g., $L$ possible single-fermion states instead of only the fully-symmetric one). In addition, we have assumed that the atoms are strongly localized, $\sigma\ll a$ (Fig. \ref{fig:lattice}). Taking into consideration the finite width of the wave packet would lead to another kind of disorder, this time associated to the interaction parameter $\beta$.

As we have pointed out, the main problem one has to face in this system is the limited lifetime of the Rydberg states, which is in the order of several microseconds. One could think of preparing a parallel system to the one described in this work but using polar molecules \cite{Micheli06,Micheli07}, to overcome this lifetime limitation.
An interesting extension is also the investigation of the system in two-dimensional geometries, e.g., triangular or square lattices, as well as several rings disposed in concentric or cylindric configurations. In all these cases, the symmetries of the particular arrangement of the sites might give rise to new interesting many-particle states.

\begin{acknowledgments}
B.O. and  R.G.F. acknowledge the grants FIS2008--02380 (MICINN),  FQM-207 and  FQM-2445 (JA), and  B.O. the support of MEC under the program FPU.
\end{acknowledgments}

\end{document}